\documentclass[prd,reprint]{revtex4-1}
\usepackage{slashed} 
\usepackage{caption} 
\usepackage{nccmath} 
\usepackage{times}
\usepackage{amsmath} 
\usepackage{amssymb} 
\usepackage{verbatim} 
\usepackage{graphicx} 
\usepackage{color} 
\usepackage{booktabs} 
\usepackage{empheq}
\usepackage{relsize}
\usepackage{fancyref}
\usepackage{xcolor,colortbl}
\definecolor{olive}{rgb}{0.3, 0.4, .1}
\definecolor{fore}{RGB}{249,242,215}
\definecolor{back}{RGB}{51,51,51}
\definecolor{title}{RGB}{255,0,90}
\definecolor{dgreen}{rgb}{0.,0.6,0.}
\definecolor{gold}{rgb}{1.,0.84,0.}
\definecolor{JungleGreen}{cmyk}{0.99,0,0.52,0}
\definecolor{BlueGreen}{cmyk}{0.85,0,0.33,0}
\definecolor{RawSienna}{cmyk}{0,0.72,1,0.45}
\definecolor{Magenta}{cmyk}{0,1,0,0}
\definecolor{lcyan}{rgb}{0.6,1,1}
\usepackage[usenames,dvipsnames,pdf]{pstricks}
\usepackage{epsfig}
\usepackage{pst-grad} 
\usepackage{pst-plot} 
\usepackage{pstricks-add}
\usepackage[off]{auto-pst-pdf} 
 
\newcommand{\lp}{\left(} \newcommand{\rp}{\right)} 
  
\newcommand{\ls}{\left[}  \newcommand{\rs}{\right]}
\newcommand{\lv}{\left|}  \newcommand{\rv}{\right|}
  
  \newcommand{\tc}{\textcolor}
\newcommand{\cd}{\!\cdot\!}
\newcommand{\st}[1]{\slashed{#1}}
\mathchardef\mhy="2D   

\DeclareMathOperator{\Tr}{Tr}

\newcommand{\idop}{\mbox{$1 \hspace{-1.3mm}  {1}$}}
\newcommand{\hs}{\hspace{\columnwidth}}
\begin{document}
\title{Fierz relations for Volkov spinors and the simplification of Furry picture traces}
\author{A. Hartin}
\affiliation{University of Hamburg/CFEL/DESY, Luruper Chausee 147, 22761, Hamburg, Germany}
\date{\today}
\maketitle
\section{Abstract}

Transition probability calculations of strong field particle processes in the Furry picture, typically use fermion Volkov solutions. These solutions have a relatively complicated spinor due to the interaction of the electron spin with a strong external field, which in turn leads to unwieldy trace calculations. The simplification of these calculations would aid theoretical studies of strong field phenomena such as the predicted resonance behaviour of higher order Furry picture processes. Here, Fierz transformations of Volkov spinors are developed and applied to a 1st order and a 2nd order Furry picture process. Combined with symmetry properties, the techniques presented here are generally applicable and lead to considerable simplification of Furry picture analytic calculations.

\section{\label{sect:intro}Introduction}

The advent of new ultra-high intensity lasers for fundamental physics research \cite{ELI16,Vulcan06} has lead to a renewal in theoretical interest in strong field physics processes \cite{DiPiaz12}. The basic framework - inherited from the similar flurry of work spurred by the invention of the laser itself in the 1960s \cite{NikRit64a,NikRit64b,NikRit65,Ritus79} - is quantum field theory in the Furry interaction picture \cite{Furry51,Schweber62,Hartin11a}. \\

In the Furry interaction picture, the gauge field is manipulated at Lagrangian level to result in scattering processes which are non-perturbative with respect to the strong external field and perturbative with respect to the interaction between quantised fields. Scattering amplitudes can be generated in the usual fashion with the S-matrix \cite{JauRoh76,FraGitShv91} or path integral method \cite{VaiFarHot92}. \\

Unlike Feynman processes in non-external field perturbation theory, one-vertex processes in the Furry picture are possible due to the contribution of momentum from the external field. These include the one photon pair production and photon radiation processes much studied in a variety of external fields and contexts \cite{NarNikRit65,Lyulka75,Bamber99,Panek03a,HarHeiIld09,BocFlo09,MacDiP11,SeiKam11,Hartin11a,Hartin15}. Higher order Furry picture processes exhibit a resonance structure as propagators go on shell for a variety of experimentally accessible particle kinematics \cite{Oleinik67,Oleinik68,Bos79a,Bos79b}. \\

This predicted resonance structure for the Furry picture propagator is one important reason to study higher order Furry picture processes in detail. The experimental investigation of this resonance structure is now feasible and would be an important test of strong field physics theoretical predictions. Theoretical studies of higher order Furry picture phenomena that exhibit this resonance behaviour include two-vertex Compton scattering \cite{Oleinik67,Oleinik68,Roshchup96,Hartin06,Seipt12}, the trident process \cite{HuMulKei10,Ildert11,King13} and loop processes \cite{Ritus72,BecMit75,BecMit76,DiPiaz12,Dinu14}. \\

The calculation of higher order Furry picture processes is somewhat complicated by the relative complexity of the scattering amplitudes \cite{Hartin06}. In contrast, a relatively simple analytic form and a simple method of calculating these amplitudes would expedite future experimental searches for Furry picture phenomena. It is the purpose of this paper to lay the ground work for a general simplification of Furry picture scattering amplitudes using Fierz transformations and general symmetry properties. \\

Fierz transformations, which allow the rearrangement of Dirac spinors in quadrilinear occurrences, have proven useful in reducing the complexity of multi fermion interactions \cite{Fierz37,Okun16}. Generalized Fierz relations which permit any interchange of pairs of spinors in a quadrilinear occurrence have been derived \cite{NiePal04}. These relations have been applied to chiral spinors \cite{Nishi05} and spin 3/2 particles, as well as the usual spin 1/2 \cite{LiaoLiu12}. To the author's knowledge, Fierz relations haven't previously been applied to Furry interaction picture processes.\\

Attempts have been made previously to simplify the calculation of higher order Furry picture processes. Often, kinematic approximations such as non-relativistic \cite{Oleinik67} or ultra-intense field intensity \cite{NikRit64a,NikRit64b} assumptions have been made. On other occasions, scattering geometries have been restricted in order to simplify analytic forms \cite{AkhMer85}. More recently, the dressed vertex has been manipulated to extract a pseudo-tensor dependence on the external field \cite{DiPiaz13}, or by exploiting gauge invariance to cancel terms \cite{Seipt12}. \\

Traditionally, exact solutions of fermions in a plane-wave external field have been applied in Furry picture calculations \cite{Volkov35}. There are however, solutions possible in centrally symmetric (Coloumb) fields and in a combination of Coulomb and plane wave fields \cite{BagGit90}. More recently, pulsed fields appropriate for actual experiments involving ultra-intense lasers, have been applied to Furry picture processes \cite{BocFlo09,HeiSeiKam10,SeiKam11}. However, efforts to use pulsed fields as a way to inhibit Furry picture resonance behaviour, appear to be unsound \cite{Ildert11}. Here, only the Volkov solution will be considered with an extension to pulsed external fields left to further work.\\

This paper intends to do the following: both the Volkov fermion wavefunction and the  strong field propagator will be cast into alternative forms in which the action is manifest in the spinor. Then, the Fierz transformation for these so-called Volkov spinors will be determined. As a means of testing this new application of the Fierz transformation, the trace of the one vertex photon radiation process in a strong external plane wave field will be obtained and compared with literature results. Finally, the developed method will be applied, as an example, to the two vertex Compton scattering in a strong external field and the result cross checked with the Klein-Nishina trace by letting the external field strength tend to zero. \\

In terms of conventions, a metric with a (1,-1,-1,-1) signature will be used. Natural units will be utilized and the Lorenz gauge will be chosen. The Lorentz gauge requires scalar products of external field 4-momentum and 4-potential to vanish, $k.A^e=0$ which will prove useful in obtaining alternative forms for the Volkov solutions and fermion propagator, 

\begin{align}\label{eq:notn}
\text{metric: }&\quad g_{\mu\nu}=(1,-1,-1,-1) \notag\\
\text{units: }&\quad c=\hbar=4\pi\epsilon_0=1, \quad e=\sqrt{\alpha} \\
\text{gauge: }&\quad \partial^\mu A^e_\mu(k\cd x)=0 \implies k\cd A=0 \notag
\end{align}

The only notation worth mentioning is the $\pm$ superscript. In all cases it indicates a linear combination, except for the Volkov spinor, $V^\pm$ where it indicates the positive and negative energy solutions.
\section{\label{sect:fierz}Volkov solutions and Fierz transformations}

In order to write down the scattering amplitude for a Furry picture process, the Volkov wavefunction solution \cite{Volkov35}, $\psi^\text{FP}_{\text{prx}}$ of an electron of momentum $p_\mu\!=\!(\epsilon,\vec{p})$, mass m and spin r embedded in a plane wave electromagnetic field of potential $A^e_{\text{x}\mu}$ and momentum $k_\mu=(\omega,\vec{k})$ is required,

\begin{gather}\label{eq:Volkov}
 \Psi^\text{FP}_\text{prx}= n_\text{p}\,E_\text{px}\; u_{\text{pr}}\;e^{- i p\cdot x },\quad n_\text{p}=\sqrt{\mfrac{m}{2\epsilon(2\pi)^3}}\, \\
E_\text{px}\equiv\ls 1 - \mfrac{\slashed{A}^e_\text{x}\st{k}}{2(k\cd p)}\rs e^{- i\mathlarger{\int}^{k\cdot x}{\;\frac{2A^{e}_\xi\cd p - A^{e\,2}_\xi}{2k\cdot p}}d\xi}, \quad A^e_{\text{x}\mu}\equiv eA^e_\mu(k\cd x) \notag
\end{gather}

The Volkov solution can be cast into an alternative form, by extracting the external field 4-momentum $\st{k}$, as a factor. The commutation properties of the slash vector enabled by the Lorenz gauge, combined with the Dirac equation give the equivalence of the two forms,

\begin{gather}\label{eq:Volkalt}
 \Psi^\text{FP}_\text{prx}= n_p\ls\st{\Pi}_{px}+m\rs\mfrac{\st{k}}{2k\cd p}\, u_\text{pr}\,e^{- i \Delta_{\text{px}} } \notag\\
\Pi_{\text{px}\mu}\equiv p_\mu-A^e_{\text{x}\mu}+\mfrac{2A^e_\text{x}\cd p - A_\text{x}^{e\,2}}{2k\cd p}k_\mu,\notag\\
\Delta_\text{px}\equiv p\cd x+\int^{k\cdot x}\mfrac{2A^e_\xi\cd p-A_\xi^{e\, 2}}{2k\cd p}\text{d}\xi\\ 
\Pi^2_{\text{px}}=p^2=m^2,\quad \mfrac{\text{d}\Delta_{\text{px}}}{\text{d}x_\mu}=\Pi_{\text{px}\mu}+A^e_{\text{x}\mu} \notag
\end{gather}

This alternative form of the Volkov solution expresses the spinor in terms of the classical kinematic momentum of the electron in the external plane wave field, which in turn bears a close relationship to the Hamilton-Jacobi action expressed in the phase. For shorthand, the spinor of the alternative form of the Volkov solution can be referred to as a Volkov spinor $V_\text{prx}$,

\begin{gather}\label{eq:volkspinor}
V_{\text{prx}}\equiv \ls\st{\Pi}_{px}+m\rs\mfrac{\st{k}}{2k\cd p}\, u_\text{pr}
\end{gather}

Armed with this alternative Volkov solution and its Volkov spinor, Fierz transformations can be introduced and then applied to the simplest one vertex Furry picture process. \\

The Fierz transformation allows the exchange of Dirac spinors within a larger product of spinors and Dirac gamma matrices arising in a scattering amplitude \cite{Ticc99}. A basis set containing five groups, scalar, vector, tensor, axial and pseudo-scalar (S,V,T,A,PS), consisting of sixteen matrices in total, is constructed,

\begin{gather}
\Gamma^i_{\text{J}}\in \{\idop,\gamma^\alpha,\sigma^{\alpha\beta},\gamma^\alpha\gamma^5,\gamma^5\}\equiv\{\Gamma_\text{S},\Gamma_\text{V},\Gamma_\text{T},\Gamma_\text{A},\Gamma_\text{PS}\} \notag\\ \alpha,\beta\in(1,2,3,4), \quad \beta>\alpha,\quad \sigma^{\alpha\beta}=\mfrac{i}{2}(\gamma^{\alpha}\gamma^{\beta}-\gamma^{\beta}\gamma^{\alpha})
\end{gather} 

The Fierz transformation is, at heart, a relationship between the matrix components (denoted by subscripts $a,b,c,d$) of the basis set under suitable contractions. The relevant properties of the basis set are the closure relation expressing orthogonality of matrix components under contraction and completeness relations \cite{NiePal04,Okun16},

\begin{gather}
\sum_\text{J}(\Gamma^i_{\text{J}})^d_a\;(\Gamma_{i\,\text{J}})^b_c=4\delta^b_a\;\delta^d_c \notag\\
\Gamma^i_{\text{I}}\Gamma^j_{\text{J}}\propto \Gamma^k_{\text{K}}, \quad M=\sum_\text{J}\mfrac{1}{4}\Tr\!\ls \Gamma_\text{Jj}M\rs\, \Gamma^\text{j}_\text{J}
\label{eq:ortho}\end{gather}

By use of these relations, Dirac spinors, $u$ within a quadrilinear (indicated by overbraces) can be swapped according to,

\begin{gather}
\ls\bar u_\text{fr}\,\Gamma^j_{\text{J}}\,u_\text{is}\,\bar u_{\text{is}'}\,\Gamma_{\text{j\,J}}u_{\text{fr}'}\rs \hs \notag \\
= \ls (\bar u_\text{fr})^a(\Gamma^j_{\text{J}})^b_a\,(u_\text{is})_b \,(\bar u_{\text{is}'})^c(\Gamma_{\text{j\,J}})^d_c\,(u_{\text{fr}'})_d \rs  \notag\\
=\sum_{\text{K}} F_{\text{JK}}\,\ls(\bar u_\text{fr})^a(\Gamma^k_{\text{K}})^d_a\,\overbrace{\tc{red}{(u_\text{is})_b}} \,(\bar u_{\text{is}'})^c(\Gamma_{\text{kK}})^b_c\,\overbrace{\tc{red}{(u_{\text{fr}'})_d}} \rs \\
=\sum_{\text{K}} F_{\text{JK}}\!\ls \bar u_\text{fr}\,\Gamma^k_{\text{K}}\,u_{\text{fr}'} \rs\ls \bar u_{\text{is}'}\,\Gamma_{\text{k\,K}}\,u_\text{is} \rs \notag \\
\text{where}\quad F_\text{JK}=\mfrac{1}{16}\Tr\ls \Gamma^j_{\text{J}} \Gamma^k_{\text{K}}\Gamma_{\text{j\,J}}\Gamma_{\text{k\,K}}\rs \notag
\label{eq:fierza}\end{gather}

The Fierz transformation matrix $F_{\text{JK}}$, is obtained from a trace over the four basis matrices appearing in the transformation and contains $5\times 5$ entries spanning the five basis groups \cite{NiePal04}. The first indice J, refers to the rows and K, the columns of the matrix,

\begin{gather}
F_{\text{JK}}=\mfrac{1}{4}
\begin{pmatrix}
1 & 1 & \frac{1}{2} & -1 & 1 \\
4 &-2 & 0  & -2 &-4 \\
12& 0 &-2 & 0 & 12 \\
-4 &-2 & 0  & -2 & 4 \\
1 &-1 & \frac{1}{2}  & 1 & 1
\end{pmatrix}
\end{gather}

The derivation of the Fierz transformation of equation 7 emphasises that the transformation involves a swap of spinors on a component-by-component basis. Since that is the case, then the spinor can be redefined ($u \rightarrow w\equiv \gamma_\mu u$) so that its components include the components of its neighbours,

\begin{gather}
\ls \bar u_\text{fr}\,\gamma^\mu\,u_\text{is}\,\bar u_{\text{is}'}\,\gamma_\mu u_{\text{fr}'} \rs \equiv \ls\bar w_\text{fr} \,\idop\,\overbrace{\tc{red}{u_\text{is}}}\,\bar u_{\text{is}'}\,\idop\, \overbrace{\tc{red}{w_{\text{fr}'}}} \rs \hs\notag\\
 =\sum_{\text{K}} F_{\text{JK}}\,\ls\bar w_\text{fr}\,\Gamma^k_{\text{K}}\,w_{\text{fr}'}\rs\ls\bar u_{\text{is}'}\,\Gamma_{\text{k\,K}}\,u_\text{is}\rs \notag\\
 \equiv \sum_{\text{K}} F_{\text{K}}\,\ls\bar u_\text{fr}\,\gamma^\mu\Gamma^k_{\text{K}}\gamma_\mu\,u_{\text{fr}'}\rs\ls\bar u_{\text{is}'}\,\Gamma_{\text{k\,K}}\,u_\text{is}\rs
\label{eq:fierzb}\end{gather}

In this way, each subsequent row of the Fierz transformation matrix can be obtained from its scalar row simply by including the basis vector in a redefinition of the spinor \cite{Okun16}. So, for instance, the vector row of the transformation matrix $F_{\text{VK}}$ is obtained by contracting over the basis vector associated with the scalar row $F_{\text{SK}}$,

\begin{gather}
F_{\text{SK}}\,\gamma^\mu \Gamma_{\text{K}} \gamma_\mu \rightarrow F_{\text{VK}}\,\Gamma_{\text{K}}\\[4pt]
F_{\text{SK}}=\begin{pmatrix} 1&1&\frac{1}{2}&-1&1 \end{pmatrix}, \quad F_{\text{VK}}=\begin{pmatrix} 4&-2&0&-2&-4 \end{pmatrix} \notag
\end{gather}

Now, the extension of Fierz transformations to Volkov spinors should be obvious. The components of the Dirac spinor can simply be redefined to be those of the Volkov spinor, $V$ (and a product with additional, $\Gamma_{\text{J}}$ matrices if desirable).  The Fierz transformation for Volkov spinors carries through virtually unchanged from the Fierz transformation for Dirac spinors, 

\begin{gather}\label{eq:volkfierz}
\ls\bar V_\text{frx}\,\Gamma^j_{\text{J}}\,\overbrace{\tc{red}{V_\text{is}}}\,\bar V_{\text{is}'\text{x}'}\,\Gamma_{\text{j\,J}}\overbrace{\tc{red}{V_{\text{fr}'\text{x}'}}}\rs \hs\notag\\
=\sum_{\text{K}} F_{\text{JK}}\,\ls\bar V_\text{frx}\,\Gamma^k_{\text{K}}\,V_{\text{fr}'\text{x}'}\rs\ls\bar V_{\text{is}'\text{x}'}\,\Gamma_{\text{k\,K}}\,V_\text{isx}\rs
\end{gather}

\section{\label{sect:hics}One-vertex Furry picture photon radiation}

\begin{figure}[h] 
\centerline{\includegraphics[width=0.5\columnwidth]{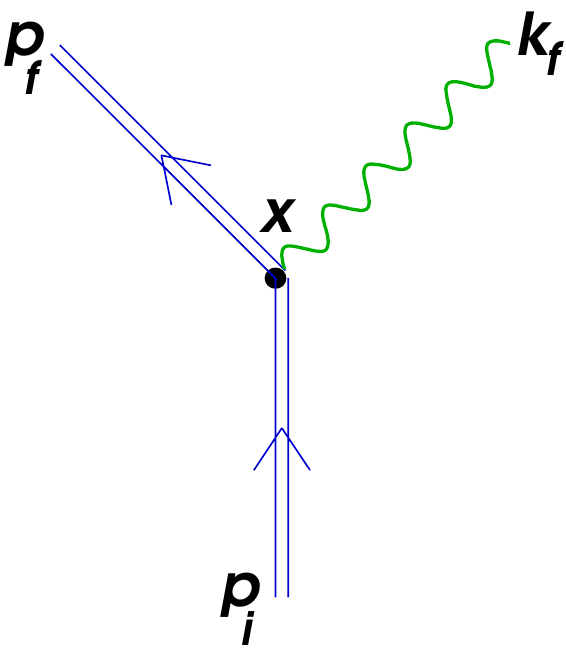}}
\caption{\bf Furry picture, one-vertex photon radiation.}\label{fig:vert}
\end{figure}

The Fierz transformation for Volkov spinors is to be applied first to the Furry picture transition probability for photon radiation from an electron in a strong electromagnetic field - the high intensity Compton scattering (HICS) process (figure \ref{fig:vert}). Here, it will not be necessary to choose a specific form for the external field. In order to cross-check the end result, an expression for the transition probability trace in a general plane-wave field, obtained by conventional methods \cite{Hartin11a} is used as a comparison. \\

The Furry picture scattering amplitude for the radiation of a photon ($k_\text{f},\st{A}_{\text{fx}}$) from an electron ($p_\text{i},\psi^\text{FP}_{\text{isx}}$) can be written down with the aid of the Feynman diagram of figure \ref{fig:vert}. Double lines indicate the Volkov wavefunction,

\begin{gather}
M^\text{HICS}_\text{f\,i}=-ie\int d^4x\; \bar \psi^\text{FP}_{\text{frx}}\,\st{A}_\text{fx} \,\psi^\text{FP}_\text{isx}
\end{gather} 

In order to concentrate on the trace, the normalisations and coupling constants will be dropped, the phases and space-time integrals will be implicit ($x'$ will indicate the additional space-time integration variable after squaring the amplitude). Photon polarisations will be summed, leaving behind their gamma matrices. Spin sums will be performed after the Fierz swap. \\

The square of the scattering amplitude will yield a quadrilinear in Volkov spinors which can be transformed as,

\begin{align}\label{eq:hics}
\sum_\text{rsr's'}&[\bar V_\text{frx}\,\gamma^\mu\,\overbrace{\tc{red}{V_\text{isx}}}][\bar V_{\text{is}'\text{x}'}\,\overbrace{\tc{red}{\gamma_\mu V_{\text{fr}'\text{x}'}}}] \notag\\
&=\sum_{\text{rsr}'\text{s}'}\sum_{\text{K}} F_{\text{SK}}\,[\bar V_\text{frx}\,\gamma^\mu\Gamma^k_{\text{K}}\gamma_\mu\,V_{\text{fr}'\text{x}'}][\bar V_{\text{is}'\text{x}'}\,\Gamma_{\text{k\,K}}\,V_\text{isx}] \notag\\
&=\sum_{\text{K}} F_{\text{SK}}\,\Tr\!\ls\bar V_\text{fx}\,\gamma^\mu\Gamma^k_{\text{K}}\gamma_\mu\,V_{\text{fx}'}\rs \Tr\!\ls\bar V_{\text{ix}'}\,\Gamma_{\text{k\,K}}\,V_\text{ix}\rs
\end{align}

After spin sums are performed, the Fierz transformation results in products of traces of Volkov spinor pairs subtending a Dirac basis group, $\Gamma_\text{K}$. The Volkov spinor pair combination appears repeatedly in this and higher order processes, suggesting that it should have its own symbol, $Q$. Taking the contraction over Dirac basis groups for granted (i.e. letting the indice $k$ be implicit and allowing the contraction $\gamma^\mu \Gamma_\text{K}\gamma_\mu$ to automatically select the vector row of the Fierz transformation matrix, $F_\text{VK}$), 

\begin{align}\label{eq:Qdef}
\Tr\!\ls\bar V_\text{fx} \Gamma_\text{K} V_{\text{fx}'}\rs &=\Tr\!\ls\mfrac{\st{k}\lp\st{\Pi}_\text{fx}+m\rp \Gamma_\text{K}\lp\st{\Pi}_{\text{fx}'}+m\rp\st{k} }{(2k\cd p_\text{f})^2}(\st{p}+m)\rs \notag\\
&\equiv \Tr\!\ls Q^\text{K}_{\text{fxx}'} \,(\st{p}+m)\rs ,\quad \text{K}\in(\text{S,V,T,A,PS})
\end{align}

The Fierz indice $K$ is a general member of the Fierz basis groups. However, the contraction that selects the vector row of the Fierz transformation matrix, automatically sets tensor contributions to zero. Additionally, the pseudo-scalar group drops out since there are insufficient gamma matrices, combined with $\gamma^5$ in the trace, to give a non-zero result. \\

Of the surviving Fierz groups (S,V,A), the symmetry of the trace under cyclic permutation and reversal allows the $Q^\text{K}_{\text{fxx}'}$ structure to appear in linear combinations $Q^{\text{K}\pm}_{\text{fxx}'}$. For the S,V groups only $Q^{+}$ appears and, for the A group which changes sign under reversal, $Q^{-}$ appears,

\begin{gather}
Q^{\text{K}\pm}_{\text{fxx}'}\equiv \mfrac{1}{2}(Q^\text{K}_{\text{fxx}'}\pm Q^\text{K}_{\text{fx}'\text{x}}),\quad \Pi^\pm_{\text{fxx}'}\equiv \mfrac{1}{2}(\Pi_\text{fx}\pm \Pi_{\text{fx}'}) \notag\\
Q^{\text{S}+}_{\text{fxx}'}= \mfrac{m\st{k}}{k\cd p}, \quad
Q^{\text{A}-}_{\text{fxx}'}= \ls\Pi^{-\alpha}_{\text{fxx}'}+\mfrac{\st{k}\st{\Pi}^{-}_{\text{fxx}'}\st{\Pi}^{+}_{\text{fxx}'}\gamma^\alpha}{2k\cd p}\rs\!\gamma^5\! \mfrac{\st{k}}{k\cd p} \notag\\
Q^{\text{V}+}_{\text{fxx}'}= \Pi^{\dag\alpha}_{\text{fxx}'}\mfrac{\st{k}}{k\cd p}, \quad \Pi^{\dag\alpha}_{\text{fxx}'}\equiv\Pi^{+\alpha}_{\text{fxx}'}+\mfrac{\st{k}}{k\cd p}\lp m^2-\Pi_\text{fx}\cd\Pi_{\text{fx}'} \rp 
\label{eq:qpmexpl}\end{gather}

The traces over explicit forms of $Q^\pm$, resulting in scalar products of $\Pi$ vectors, reduce to simpler forms by using the conservation of energy-momentum at the Furry picture (dressed) vertex (figure \ref{fig:vert}). As a result of the properties of the alternative form of the Volkov solution (equation 3), and after some mathematical work (Appendix \ref{app:momcons}),

\begin{gather} \label{eq:cofem}
\int^\infty_{-\infty} \!\text{d}x_\mu \lp \Pi_\text{fx}\!+\!k_\text{f}\!-\!\Pi_\text{ix}\rp_\mu  e^{i\lp\Delta_\text{fx}+k_f\cdot x-\Delta_\text{ix}\rp} \hs \notag\\ 
=\int^\infty_{-\infty}\! \text{d}x_\mu \mfrac{\text{d}\lp \mathsmaller{\Delta_\text{fx}+k_f\cdot x-\Delta_\text{ix}}\rp}{\text{d}x_\mu}\,e^{i\lp\Delta_\text{fx}+k_f\cdot x-\Delta_\text{ix}\rp}=0
\end{gather} 

Consequently, under the space-time integrals,

\begin{gather}\label{eq:PIsp}
(\Pi_\text{fx}-\Pi_\text{ix})^2=2m^2-2\Pi_\text{fx}\cd\Pi_\text{ix}= k_f^2=0 \notag\\
(\Pi_\text{fx}-\Pi_{\text{fx}'})^2=2m^2-2\Pi_\text{fx}\cd\Pi_{\text{fx}'}=(A^e_\text{x}-A^e_{\text{x}'})^2\equiv A_{\text{xx}'} \notag\\
4\,\Pi^{-}_{\text{ixx}'}\cd \Pi^{-}_{\text{fx}'\text{x}}= \Pi_{\text{ix}}\cd\Pi_{\text{fx}'}+\Pi_{\text{ix}'}\cd\Pi_{\text{fx}}-2m^2 \\
4\,\Pi^\dag_{\text{ixx}'}\cd \Pi^\dag_{\text{fx}'\text{x}}=\Pi_{\text{ix}}\cd\Pi_{\text{fx}'}+\Pi_{\text{ix}'}\cd\Pi_{\text{fx}}\!+\!2m^2\!+\!\ls \mfrac{k\cd p_\text{i}}{k\cd p_\text{f}}\!+\!\mfrac{k\cd p_\text{f}}{k\cd p_\text{i}} \rs\!A_{\text{xx}'} \notag
\end{gather}

Returning to the square of the scattering amplitude for the one vertex radiation process and performing the traces, there ensues

\begin{gather}
|M^\text{HICS}_\text{f\,i}|^2 \propto 16m^2+8\lp \Pi^{-}_{\text{ixx}'}\cd \Pi^{-}_{\text{fx}'\text{x}}-\Pi^\dag_{\text{ixx}'}\cd \Pi^\dag_{\text{fx}'\text{x}}\rp \notag\\
=8m^2\!-\!2\ls \mfrac{k\cd p_\text{i}}{k\cd p_\text{f}}\!+\!\mfrac{k\cd p_\text{f}}{k\cd p_\text{i}} \rs \!A_{\text{xx}'}
\label{eq:hicsresult}\end{gather}

This result for the Furry picture photon radiation trace in equation \ref{eq:hicsresult} is the same as that obtained earlier for a general plane wave external field \cite{Hartin11a}, i.e. the literature result \cite{NikRit64a,NikRit64b,NarNikRit65}. The Fierz transformation combined with energy-momentum conservation, eliminated extra terms and led to the final result much faster than with the conventional method (which is one long trace with standard Volkov solutions). \\

Next, after discussing the propagator, this method will be applied to a two vertex Furry picture process whose conventional trace calculations are difficult, even with the aid of computer programs \cite{Mertig91}.

\section{\label{sect:prop}Furry picture fermion propagator}

In order to extend the Fierz transformation of Volkov spinors to higher order processes, it is useful to transform the Furry picture fermion propagator into a sum over Volkov spinors in their alternative form. It will not be necessary to consider the propagator denominator and its unusual pole structure, beyond simply writing it down. The phase part of the propagator plays a role in energy-momentum conservation rules. \\

In its usual form, the Furry picture propagator is the free propagator subtended by Volkov $E_p$ functions \cite{Ritus72},

\begin{gather}
G^\text{FP}_\text{yx}=i\int\mfrac{\text{d}^4p}{(2\pi)^4}\;E_\text{py}\,\mfrac{\st{p}+m}{p^2-m^2+i\varepsilon}\, \bar E_\text{px}\; e^{ip\cdot (x-y)}
\end{gather}
 
An alternative form for the propagator, similar to that found for the Volkov solution in equation 3, can be obtained. This alternative form, containing $\Pi$ vectors, will once again enable simpler traces through conservation of energy-momentum at each vertex. Allowing the integration over propagator momentum and the propagator phase to be implicit (since we are only concerned with the trace calculation),

\begin{gather}\label{eq:Galt1}
G^\text{FP}_\text{yx}\propto \mfrac{\lp \st{\Pi}_\text{py}+m\rp \mfrac{\st{k}}{2k\cd p}(\st{p}+m)\mfrac{\st{k}}{2k\cd p} \lp \st{\Pi}_\text{px}+m\rp +\mfrac{\st{k}}{2k\cd p}\lp p^2-m^2 \rp}{p^2-m^2+i\varepsilon}
\end{gather}

The residual term $\frac{\st{k}}{2k\cdot p}\lp p^2-m^2 \rp$ in equation \ref{eq:Galt1}, exists because the propagator momentum is, in general, not on shell. The residual term can be eliminated with a trick. \\

A special symbol, $m_\ast$ is defined that has the property of being the virtual momentum when it is squared and the fermion mass when it appears on its own (these two rules must be applied in order),

\begin{gather}
m_\ast^2 \rightarrow p^2,\quad m_\ast\rightarrow m
\end{gather}

Using $m_\ast$, the propagator numerator can be represented with $\Pi$ vectors, as a single, symmetrical product,

\begin{gather}\label{eq:Galt2}
G^\text{FP}_\text{yx}\propto \mfrac{\lp \st{\Pi}_\text{py}+m_\ast\rp \mfrac{\st{k}}{2k\cd p}(\st{p}+m_\ast)\mfrac{\st{k}}{2k\cd p} \lp \st{\Pi}_\text{px}+m_\ast\rp }{p^2-m^2+i\varepsilon}
\end{gather}
 
In order to use this $m_\ast$ trick successfully, the appearance of the propagator in a squared amplitude must contain a separately labelled $m_\ast$ for each separate appearance. \\

The propagator numerator can now be represented by a product of positive and negative energy Volkov spinors, $V^\pm$. Cross-terms $V^\pm_\text{pty} \bar V^\mp_\text{ptx}$ vanish by orthogonality and the sum of end terms $V^\pm_\text{pty} \bar V^\pm_\text{ptx}$ give the Feynman form  of the Volkov propagator \cite{Scadron90,BerVar80a}. 

\begin{gather}\label{eq:propspinsum}
G^\text{FP}_\text{yx}=i\sum_\text{t}\int\mfrac{\text{d}^4p}{(2\pi)^4}\,\mfrac{V_\text{pty}\bar{V}_\text{ptx}}{p^2-m^2+i\varepsilon}\,  e^{i(\Delta_\text{px}-\Delta_\text{py})} \\
\text{where}\quad V_\text{pty}\equiv V^{+}_\text{pty}-V^{-}_{-\text{pty}} \notag
\end{gather}

In this final form, the Volkov propagator can be inserted into a higher order Furry picture process, and Fierz transformations for Volkov spinors can be fully exploited to reduce the trace to its simplest form.
\section{\label{sect:compt}Two-vertex, Furry picture Compton scattering}

\begin{figure}[h] 
\centerline{\includegraphics[width=0.45\textwidth]{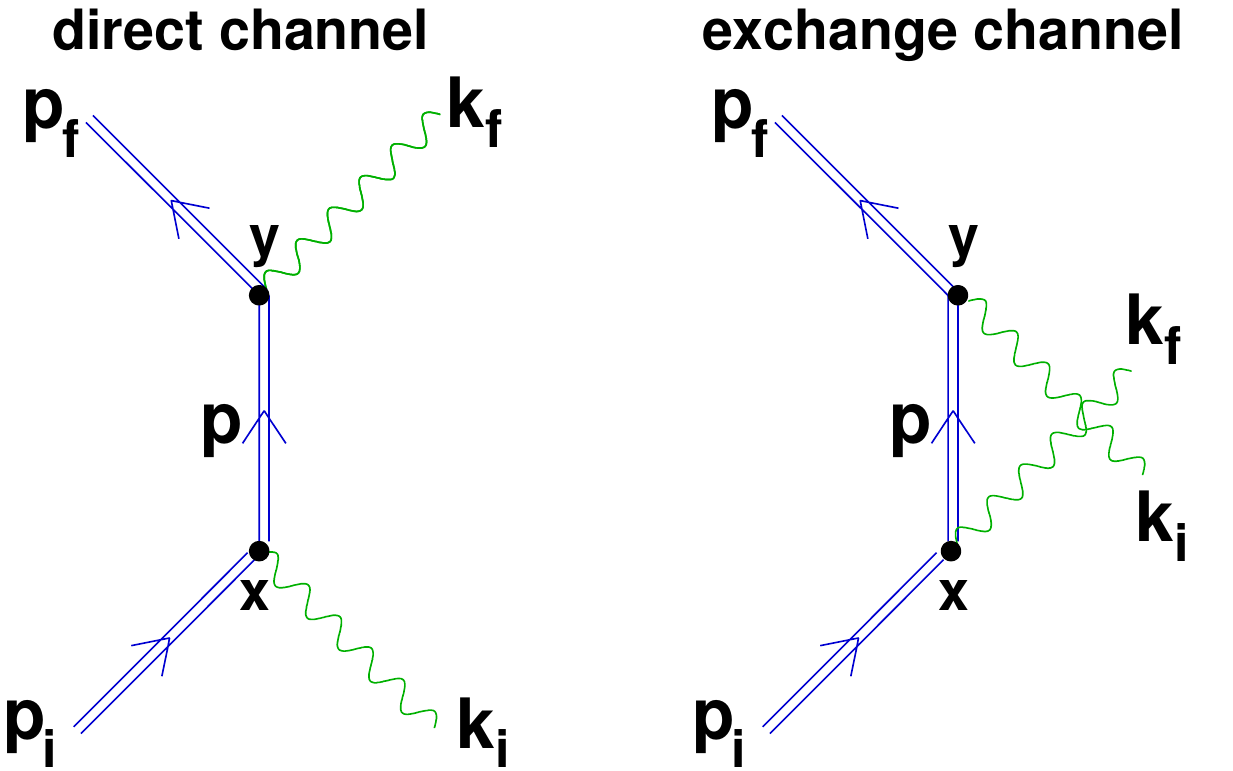}}
\caption{\bf Furry picture, two-vertex Compton scattering.}
\label{fig:compt}\end{figure} 

The method of Fierz swapping Volkov spinors will be applied to a second order Furry picture process, the stimulated Compton scattering (SCS) process (figure \ref{fig:compt}). The squared SCS amplitude will be simplified with use of the Fierz transformations of section \ref{sect:fierz} and with the alternative Volkov propagator decomposed into a spin sum (equation \ref{eq:propspinsum}). The aim is to write down the direct channel trace in a simple form. \\

The direct channel amplitude of the SCS process, after summing over photon polarizations, comprises two dressed vertices consisting of four Volkov spinors in total. Leaving the space-time integrations, spin sums, spin subscripts, propagator denominator, phases, normalisations and coupling constant implicit,

\begin{align}
M^\text{SCS}_\text{f\,i}&=-ie\int \text{d}^4\text{x}\, \text{d}^4\text{y}\;\; \bar \psi^\text{FP}_{\text{fry}}\,\st{A}_\text{fy}\, G^\text{FP}_\text{yx}\,\st{A}_\text{ix}\,\psi^\text{FP}_\text{isx} \notag\\
&\rightarrow\quad \bar V_\text{fy}\,\gamma^\mu\,V_\text{py}\bar V_\text{px}\,\gamma^\nu\,V_\text{ix}
\end{align}

Though there are twice as many spinors in the SCS amplitude as compared to the one-vertex HICS amplitude, there are correspondingly, two energy-momentum conservation rules grouped around each vertex and, possibly, additional Fierz transformations. After making two Fierz transformations (indicated by over and under braces) which produce sums over two basis groups, J,K, and carrying out the appropriate spin sums, a product of three traces of familiar $Q$ structures results,

\begin{gather}\label{eq:fierzswap}
\bar V_\text{fy}\,\gamma^\mu\,\overbrace{\tc{dgreen}{V_\text{py}}}\bar V_\text{px}\,\underbrace{\tc{red}{\gamma^\nu\,V_\text{ix}}}\bar V_{\text{ix}'}\,\gamma_\nu\,\underbrace{\tc{red}{V_{\text{px}'}}}\bar V_{\text{py}'}\,\overbrace{\tc{dgreen}{\gamma_\mu V_{\text{fy}'}}} \\[4pt]
\rightarrow \sum_\text{JK}F_\text{VJ}F_\text{VK}\!\Tr\!\ls Q^\text{J}_{\text{fy}'\text{y}} (\st{p}_f\!+\!m) \rs \!\Tr\!\ls Q^\text{K}_{\text{ixx}'} (\st{p}_i\!+\!m)\rs \notag\\
\centerdot\Tr\!\ls Q^\text{K}_{\text{pxx}'}(\st{p}_{}+m_{\ast'})Q^\text{J}_{\text{py}'\text{y}}(\st{p}+m_\ast) \rs \notag
\end{gather}

The first two traces in equation \ref{eq:fierzswap} have the same form as those obtained for the one vertex HICS process. This ensures that, once again, $Q^\pm$ combinations appear and only the (S,V,A) Fierz groups survive. The third trace of equation \ref{eq:fierzswap} can also be reordered into $Q^\pm$ combinations via a symmetry with respect to complex conjugation. \\

The integration over the space-time coordinate at each vertex, say $x$ appears with an additional $x'$ in the squared amplitude. The dependence on both these space-time coordinates appears as some function $F_{\text{xx}'}$ in the trace with some phase $P_\text{x}-P_{\text{x}'}$ as,

\begin{gather}
\int \text{d}x\,\text{d}x' \, F_{\text{xx}'}\, e^{i\lp P_\text{x}-P_{\text{x}'}\rp} 
\end{gather}

Assuming $F$ is a real valued function (and it is if the external field 4-potential $A^e$ is written as a real function), the operation of swapping $x\rightarrow x'$ results in the complex conjugate. However, since this is a squared amplitude, taking the complex conjugate must leave it unchanged. Therefore, the function $F$ can be replaced by 

\begin{gather}
F_{\text{xx}'}\rightarrow \mfrac{1}{2}\ls F_{\text{xx}'}+F_{\text{x}'\text{x}}\rs
\end{gather}

Moreover, this complex conjugate symmetry operation must be applicable at each vertex separately since there is no other way to ensure invariance with respect to overall complex conjugation in the squared amplitude of a multi-vertex process (see Appendix \ref{app:comconj}). The vertex independence of the complex conjugation symmetry allows the $Q^\pm$ linear combinations to be formed in the third trace of equation \ref{eq:fierzswap}. \\

This third trace, however, contains non-mass shell 4-momenta, $p$ with associated "alternative" masses (labelled separately), $m_\ast,m_{\ast'}$. Consequently, the $Q^\pm$ combinations for non-mass shell momenta, contain extra terms which reduce to their previous form (equation \ref{eq:qpmexpl}) once the mass shell condition is applied,

\begin{gather}
Q^{S+}_{\text{pxx}'}\!=\! \mfrac{(m_\ast\!+\!m_{\ast'})\st{k}}{2k\cd p}, \; Q^{V+}_{\text{pxx}'}\!=\! \ls\Pi^{\dag\alpha}_{\text{fxx}'}+\mfrac{(m_{\ast'}\!-\!m_{\ast})\st{k}\st{\Pi}^{+}_{\text{pxx}'}\gamma^\alpha}{4k\cd p}\rs\!\mfrac{\st{k}}{k\cd p} \notag \\
Q^{A-}_{\text{pxx}'}= \ls\Pi^{-\alpha}_{\text{pxx}'}+\mfrac{(m_{\ast}-m_{\ast'})k^\alpha\st{\Pi}^{-}_{\text{pxx}'}+\st{k}\st{\Pi}^{-}_{\text{pxx}'}\st{\Pi}^{+}_{\text{pxx}'}\gamma^\alpha}{2k\cd p}\rs\!\gamma^5\! \mfrac{\st{k}}{k\cd p} 
\label{eq:qpmexpl2}\end{gather}

Returning to the Fierz swapped, squared amplitude of the SCS direct channel (equation \ref{eq:fierzswap}), applying the explicit forms of the $Q^\pm$ combinations and performing the traces, a compact form is obtained,

\begin{gather}
|M^\text{SCS}_\text{f\,i}|^2 \propto 16(p^2-m^2)\!\ls 4m^2\!-\!X^\text{fp}_{\text{y}'\text{y}}\cd X^\text{ip}_{\text{xx}'}\rs+32 Y^\text{fp}_{\text{y}'\text{y}}\cd Y^\text{ip}_{\text{xx}'} \notag\\
\text{where}\quad X^\text{ip}_{\text{xx}'}\equiv \lp \Pi^\dag_{\text{ixx}'} -\mfrac{k\cd p_i}{k\cd p}\Pi^{-}_{\text{pxx}'}\rp \\
Y^\text{ip}_{\text{xx}'}\equiv \Pi^\dag_{\text{ixx}'}\cd \Pi^\dag_{\text{pxx}'}-\Pi^{-}_{\text{ixx}'}\cd \Pi^{-}_{\text{pxx}'}-2m^2 \notag
\end{gather}

The scalar products of $\Pi$ vectors can be reduced to simple forms, once again, by using conservation of energy-momentum at each vertex. In the limit of vanishing field strength, the $\Pi^{-}$ vectors go to zero, $\Pi^\dag$ goes to $\Pi^{+}$, $\Pi^{+}$ vectors lose their spatial dependence and reduce to simple 4-momenta, and the Klein-Nishina direct channel trace \cite{ManSha84} is easily obtained. \\

The above methods apply readily to the exchange channel since only the internal momentum differs between the direct ($p\equiv p_\text{d}$) and the exchange ($p\equiv p_\text{e}$) channels. \\

 The interference terms (direct $\times$ exchange amplitudes) however, have a different internal structure in the amplitude with respect to contracted gamma matrices. In that case, the Fierz swaps can be made between slightly different groups of Volkov spinors and gamma matrices, and the same structure results.

\begin{align}
&\bar V_\text{fy}\,\overbrace{\tc{dgreen}{\gamma^\mu\,V_\text{dy}}}\bar V_\text{dx}\,\gamma^\nu\,V_\text{ix}\bar V_{\text{ix}'}\,\gamma_\mu\,\overbrace{\tc{dgreen}{V_{\text{ex}'}\bar V_{\text{ey}'}\,\gamma_\nu V_{\text{fy}'}}} \notag\\[4pt]
&\rightarrow\ls V_\text{ix}\bar V_{\text{ix}'}\,\gamma_\mu\,\Gamma^\text{K}\gamma^\mu\rs\bar V_\text{fy}\Gamma^\text{K}V_{\text{ex}'}\bar V_{\text{ey}'}\,\underbrace{\tc{red}{\gamma_\nu V_{\text{fy}'}}}\bar V_\text{dx}\,\gamma^\nu\underbrace{\tc{red}{V_\text{dy}}} \notag\\
&\rightarrow\ls V_{\text{ix}}\bar V_{\text{ix}'}\,\gamma_\mu\,\Gamma^\text{K}\gamma^\mu\rs\ls V_{\text{fy}'}\bar V_{\text{fy}}\,\gamma_\nu\,\Gamma^\text{J}\gamma^\nu\rs \ls\bar V_\text{dx}\Gamma^\text{K}V_{\text{ex}'}\bar V_{\text{ey}'}\Gamma^\text{J}V_\text{dy}\rs &
\label{eq:fierzswapint}\end{align}

Both the direct and exchange internal momenta appear in the third group of Fierz swapped Volkov spinors in the interference terms, so a further generalisation of the $Q^\pm$ vectors, the most general one possible, is necessary,

\begin{gather}
Q^{\text{K}}_{\text{dxex}'}\equiv\bar V_\text{dx}\Gamma^\text{K}V_{\text{ex}'}, \quad Q^{\text{K}\pm}_{\text{dxex}'}\equiv\mfrac{1}{2}\lp Q^{\text{K}}_{\text{dxex}'}\pm Q^{\text{K}}_{\text{dx}'\text{ex}}\rp \notag\\
Q^{S+}_{\text{dxex}'}= \lp\!\mfrac{m_\ast-\st{\Pi}^{+}_{\text{dxx}'}}{2k\cd p_\text{d}}+\mfrac{m_{\ast'}+\st{\Pi}^{+}_{\text{exx}'}}{2k\cd p_\text{e}}\!\rp\st{k} \\
Q^{V+}_{\text{dxex}'}\!=\! \ls\Pi^{+\alpha}_{\text{exx}'}\!+\!\mfrac{2m_\ast m_{\ast'}k^\alpha\!+\!\st{k}(\st{\Pi}^{-}_{\text{dxx}'}\st{\Pi}^{-}_{\text{exx}'}\!-\!\st{\Pi}^{+}_{\text{dxx}'}\st{\Pi}^{+}_{\text{exx}'} \!+\! m_{\ast'}\st{\Pi}^{+}_{\text{dxx}'}\!-\!m_{\ast}\st{\Pi}^{+}_{\text{exx}'})\gamma^\alpha}{4k\cd p_\text{d}}\rs \!\mfrac{\st{k}}{k\cd p_\text{e}} \notag \\
Q^{A-}_{\text{dxex}'}\!=\! \ls\Pi^{-\alpha}_{\text{exx}'}+\!\mfrac{2(m_\ast\st{\Pi}^{-}_{\text{exx}'} \!-\!m_{\ast'}\st{\Pi}^{-}_{\text{exx}'})k^\alpha \!+\!\st{k}(\st{\Pi}^{-}_{\text{dxx}'}\st{\Pi}^{+}_{\text{exx}'}\!-\!\st{\Pi}^{+}_{\text{dxx}'}\st{\Pi}^{-}_{\text{dxx}'})\gamma^\alpha}{4k\cd p_\text{d}}\rs \!\gamma^5 \!\mfrac{\st{k}}{k\cd p_\text{e}} \notag
\label{eq:qpmexpl3}\end{gather} 

The above $Q^\pm$ vectors reduce to those of previous incarnations (equations \ref{eq:qpmexpl} and \ref{eq:qpmexpl2}) when the internal momenta $p_\text{d}, p_\text{e}$ are the same, and/or are on the mass shell.

\section{Conclusion}

In this paper, a general method for simplifying scattering amplitudes for Furry picture particle processes has been developed, demonstrated and cross-checked in two cases. By making Fierz swaps of Volkov spinors, and with application of general energy-momentum conservation and complex conjugation symmetries, the one-vertex photon radiation (HICS) trace result was obtained with a few lines of algebra. The two vertex Compton scattering (SCS), a traditionally difficult process to analytically calculate, was similarly obtained in compact form. This method can be referred to as the FTVS (Fierz transformation of Volkov Spinors).\\

In preparation, the traditional Volkov solutions for fermions embedded in an external, plane wave, electromagnetic field were cast into an alternative form in which the action of the electron embedded in the external field is manifest in the Volkov spinor. The method of Fierz transformations for Dirac spinor quadrilinears was extended to Volkov spinors by taking note that the Fierz transformation is a relation between a basis set of five groups with spinors swapping component-wise. \\

When applying this Fierz swapping of Volkov spinors to actual squared scattering amplitudes, repeated patterns emerged. A structure of two Volkov spinors subtending a Dirac basis group was given its own symbol, $Q^\text{K}$ with the superscript, K ranging over the Dirac matrix basis groups. When spin sums are performed, $Q^\text{K}$ appears only once or twice in a trace.  \\

Because of contraction over Dirac gamma matrices, arising from polarisation sums, each appearance of a Fierz transformation matrix is via its vector row. This, combined with the limited size of the resultant traces, ensures that only scalar, vector and axial basis groups appear. The physical explanation for the (S,V,A) structure is that, besides the photon vector current interaction, there is also a coupling to the axial vector current provided by the plane wave external field \cite{DiPiaz13}. Indeed, as the external field vanishes it can be seen analytically that axial parts of the trace drop out. \\

\begin{figure}[h] 
\centerline{\includegraphics[width=0.45\columnwidth]{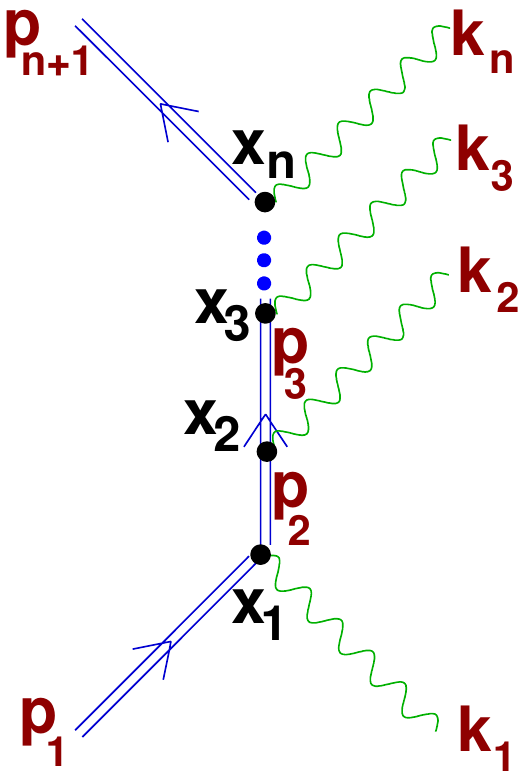}}
\caption{\bf An n-vertex Furry picture process.}\label{fig:nvert}
\end{figure} 

Symmetry properties were applied. Invariance of the squared amplitude under complex conjugation at each vertex, combined with reversal symmetry of the trace, allowed only three linear combinations, $Q^\text{S+},Q^\text{V+},Q^\text{A-}$. Each of these linear combinations in turn led to simple scalar products of $\Pi$ vectors. Finally, conservation of energy-momentum was used to simplify $\Pi$ scalar products further. The procedure was verified by obtaining known results in one-vertex and two-vertex processes. \\

The FTVS method described above should be applicable more generally. For instance, in an n-vertex Furry picture process (figure \ref{fig:nvert}), a quick "power counting" establishes the expected structures. \\

An amplitude with n vertices will give rise to 2n Volkov spinor pairs (i.e. 2n Q structures) when the amplitude is squared. In turn, n Fierz swaps will give rise to n traces each containing one $Q^\text{K}$ structure. The remaining n Q structures will appear in a single trace. There will be n sets of complex conjugation symmetries to produce $Q^\pm$ linear combinations, and n energy-momentum conservation requirements that will ensure simple forms for the scalar products of $\Pi$ vectors. \\

FTVS can be applied to Furry picture helicity amplitudes when specific spin/polarisation states are the aim of the calculation. In that case, the Dirac basis set can be reformed as a helicity basis set, which is more efficient for Fierz transformations of helicity amplitudes \cite{Nishi05}. \\

The extension of FTVS to pulsed or otherwise non-plane wave external fields \cite{SeiKam11,BagGit90} should present little difficulty. The Fierz transformation and exploited symmetries are general. In terms of simple results it will be useful to express the action of the electron in the spinor of any non-Volkov solution. \\

Other Furry picture Feynman diagrams, those containing internal photon lines and/or loops, should also be amenable to FTVS. Loops are in any case related to tree level diagrams via cutting rules \cite{Cutkos60}, so it is expected that FTVS can seamlessly be applied. The $Q$ structures are simply an expression of the Volkov current through a dressed vertex, which is the basis of a Furry picture perturbation expansion. \\

It is expected that FTVS will prove useful for a range of higher order Furry picture calculations. The author is particularly interested in detailed consideration of the two-vertex SCS and associated processes. To that end, the FTVS will be combined with a detailed treatment of Volkov phases and propagator denominators in future work. It is also anticipated that FTVS will prove useful, for example, in a Bloch-Nordsieck-type proof \cite{BloNor37,PesSch95} of divergence cancellation and for the inclusion of one-particle irreducible diagrams in complete Furry picture propagators. 

\begin{acknowledgments}
The author would like to acknowledge funding from the Partnership of DESY and Hamburg University (PIER) for the seed project, PIF-2016-53.
\end{acknowledgments}

\appendix\section{Energy-momentum conservation with Volkov solutions under space-time integration}\label{app:momcons}

Energy-momentum conservation in strong field processes requires that the momentum flow through a dressed vertex be zero,

\begin{gather} \label{appeq:cofem}
\int^\infty_{-\infty} \!\text{d}x_\mu \lp \Pi_\text{fx}\!+\!k_\text{f}\!-\!\Pi_\text{ix}\rp_\mu  e^{i\lp\Delta_\text{fx}+k_f\cdot x-\Delta_\text{ix}\rp} =0
\end{gather}

Equation \ref{appeq:cofem} has been demonstrated to be mathematically true in the context of linearly polarised laser fields \cite{NikRit64a,Ildert11,MacDiP11}. The solution presented here, for a general plane wave 4-potential $A^e_\mu(k\cd x)$, is based partly on \cite{Ritus79} and makes use of the Fourier expansion of a function $f$ of arbitrary period $L$ \cite{Hartin11a},

\begin{gather}\label{appeq:FTarbitrary}
f(k\cd x)\equiv\sum\limits_{r=-\infty}^{\infty}\int_{-\pi L}^{\pi L} \mfrac{d\phi}{2\pi L}
\,e^{i\frac{r}{L}\ls \phi-k\cdot x\rs} f(\phi) 
\end{gather}

The relationship between the classical momentum and the Hamilton-Jacobi action of the electron in the external field (equation 3) simplifies the integrand of \ref{appeq:cofem}, and the non-linear dependence on the space-time variable, say $f(k\cd x)$, can be made linear by using the Fourier expansion (\ref{appeq:FTarbitrary}),

\begin{gather} \label{eq:cofem}
\int^\infty_{-\infty} \!\text{d}x_\mu \lp \Pi_\text{fx}\!+\!k_\text{f}\!-\!\Pi_\text{ix}\rp_\mu  e^{i\lp\Delta_\text{fx}+k_f\cdot x-\Delta_\text{ix}\rp} \hs \notag\\ 
=\int^\infty_{-\infty}\! \text{d}x_\mu \mfrac{\text{d}}{\text{d}x_\mu}\!\ls\,e^{i\lp\Delta_\text{fx}+k_f\cdot x-\Delta_\text{ix}\rp}\rs \notag\\
=\sum\limits_{r=-\infty}^{\infty}\int_{-\pi L}^{\pi L} \mfrac{d\phi}{2\pi L}\,e^{i\frac{r}{L}\phi}f(\phi)\int^\infty_{-\infty}\! \text{d}x_\mu \mfrac{\text{d}}{\text{d}x_\mu}e^{i\lp p_\text{f}+k_f-p_\text{i}-\frac{r}{L}k\rp\cdot x}
\end{gather} 

The derivative and integration with respect to space-time, result in a product of a delta function and its argument, and the identity (equation \ref{appeq:cofem}) is proved,

\begin{gather}
\int^\infty_{-\infty}\! \mfrac{\text{d}x_\mu}{(2\pi)^4} \,\mfrac{\text{d}}{\text{d}x_\mu}\,e^{i\lp p_\text{f}+k_f-p_\text{i}-\frac{r}{L}k\rp\cdot x}= \hs\notag\\
\lp p_\text{f}+k_f-p_\text{i}-\mfrac{r}{L}k\rp_{\!\mu} \delta^4\!\lp p_\text{f}+k_f-p_\text{i}-\mfrac{r}{L}k\rp =0
\end{gather}

\section{\label{app:comconj}Complex conjugation symmetry in squared amplitudes}

The configuration-space amplitude of a general, two-vertex Furry picture particle process requires integration over two space-time coordinates $x,y$, with additional $x',y'$ coordinates (and integrations) appearing when the amplitude is squared. Symmetries of the integrand of these space-time integrals under the operation of complex conjugation will be considered. \\

In a Furry picture process, there is a space-time dependence in the spinor, say $f_{\text{y}},g_{\text{x}}$ and in the phase, $P_\text{x},Q_\text{y}$. The dependence in the squared amplitude then appears as a product of sub-amplitudes and their conjugates,

\begin{gather}
\lv M_\text{fi}\rv^2 \!\propto\! \int\! \text{d}x\,\text{d}x'\text{d}y\,\text{d}y' \Tr\! \ls f_{\text{y}}\,g_{\text{x}}\,\bar g_{\text{x}'} \bar f_{\text{y}'} \rs\!e^{i\lp P_\text{x}-P_{\text{x}'}\rp + i\lp Q_\text{y}-Q_{\text{y}'}\rp}
\end{gather}

Using the symmetry properties of the trace, the dependence on space-time coordinate conjugate pairs $x,x'$ and $y,y'$ is separable ($F_{\text{y}'\text{y}}\equiv \bar f_{\text{y}'}f_\text{y}$, $G_{\text{xx}'}\equiv g_\text{x}\bar g_{\text{x}'}$),

\begin{gather}
\int\! \text{d}x\,\text{d}x'\text{d}y\,\text{d}y' \Tr\! \ls F_{\text{y}'\text{y}}\,e^{i\lp Q_\text{y}-Q_{\text{y}'}\rp}\,G_{\text{xx}'}\,e^{i\lp P_\text{x}-P_{\text{x}'}\rp} \rs\
\end{gather}

The integrand dependence on each conjugate pair of space-time coordinates appears as a squared function times an imaginary phase. Therefore, the operation of swapping each space-time conjugate pair, $x\leftrightarrow x'$ or $y\leftrightarrow y'$, must leave the integrand real valued, assuming the functions $F,G,P$ and $Q$ are real valued. In other words, the imaginary parts must vanish,

\begin{gather}
\int \text{d}y\,\text{d}y'\; F_{\text{y}'\text{y}}\,\sin\!\lp Q_\text{y}-Q_{\text{y}'}\rp = 0
\label{eq:app1}\end{gather}

Writing the functions, $F$ and $G$, in two terms which are odd and even with respect to swapping the space-time coordinates,

\begin{gather}
F_{\text{y}'\text{y}}=F^{+}_{\text{y}'\text{y}}+F^{-}_{\text{y}'\text{y}}, \quad F^\pm_{\text{y}'\text{y}}\equiv \mfrac{1}{2}\lp F_{\text{y}'\text{y}} \pm F_{\text{yy}'} \rp
\end{gather}

the requirement of invariance under complex conjugation, of the integrand of each squared sub-amplitude, ensures

\begin{gather}
 F_{\text{y}'\text{y}}\rightarrow F^{+}_{\text{y}'\text{y}},\; G_{\text{xx}'}\rightarrow G^{+}_{\text{xx}'}
\end{gather}

\bibliographystyle{unsrtnat}
\bibliography{/home/hartin/Physics_Research/mypapers/hartin_bibliography}

\end{document}